\documentclass[12pt,a4paper]{article}
\usepackage[utf8]{inputenc}
\usepackage{amsmath}
\usepackage{amsfonts}
\usepackage{amssymb}

\usepackage{hyperref}
\usepackage{graphicx}
\usepackage{lmodern}
\usepackage[left=2cm,right=2cm,top=2cm,bottom=2cm]{geometry}
\title{Constructing Inverse Scattering Potentials \\for $\alpha-\alpha$ System using Reference Potential Approach }
\author{O. S. K. S. Sastri, Arushi Sharma and Ayushi Awasthi}
\begin{document}
\maketitle
\begin{abstract}
\begin{description}
\item[Background]
An accurate way to incorporate long range Coulomb interaction alongside short-range nuclear interaction has been a challenge for theoretical physicists.
\item[Purpose]
In this paper, we propose a methodology based on the reference potential approach for constructing inverse potentials of alpha-alpha scattering.
\item[Methods]
Two smoothly joined Morse potentials, regular for short-range nuclear interaction and inverted for long range Coulomb, are used in tandem as a reference potential in the phase function method to obtain the scattering phase shifts for the S, D and G states of alpha-alpha scattering. The model parameters are optimized by choosing to minimize the mean absolute percentage error between the obtained and experimental scattering phase shift values.
\item[Results]
The constructed inverse potentials for S, D and G states have resulted in mean absolute percentage errors of 0.8, 0.5, and 0.4 respectively. The obtained resonances for D and G states closely match the experimental ones.
\item[Conclusion]
The reference potential approach using a combination of smoothly joined Morse functions is successful in accurately accounting for the Coulomb interaction between charged particles in nuclear scattering studies.
\end{description}

\end{abstract} 

\textit{Introduction:} One of the most difficult problems for theoretical physicists is acquiring an accurate understanding of scattering involving a charged projectile and a charged target due to the indefinitely long range of Coulomb interaction\cite{1}. Inadvertently, microscopic theories such as phenomenological potential models used in R-matrix\cite{2}, complex scaling method\cite{3}, phase function method\cite{4, 5}, and others face mathematical difficulties in including Coulomb effect in a rigorous manner. As a result, they use screened Coulomb potentials to simulate the often observed condition, in which an isolated charge is generally surrounded by residual particles due to polarization. Taylor\cite{1} presented a comprehensive method of incorporating Coulomb scattering by considering
\begin{equation}
V_c^\rho (r) = \frac{\gamma}{r}\alpha^\rho (r) 
\end{equation}
where the screening function $\alpha^\rho$ for a given $\rho$ must go to zero as r tends to $\infty$ and must approach 1 as $\rho$ tends to $\infty$ with r fixed. As long as $\rho$ is very large, such a potential meets the conditions of scattering theory and produces findings that are independent of both properties of screened potential, their nature/shape and screening radius.\\
To solve this crucial issue, we focused on $\alpha-\alpha$ scattering in our article.
Ali-Bodmer\cite{6}, Buck\cite{7} and Odsuren\cite{3} utilise an \textit{erf()} function to model the Coulomb interaction, 
\begin{equation}
V_{c}=\frac{4e^2}{r}erf(\beta r),  \text{and}~~\beta=\frac{\sqrt{3}}{2R_\alpha}
\label{error}
\end{equation}
The interaction resulting from this statement is referred to as an enhanced Coulomb interaction.This results from the fact that $\alpha$-particles have a finite size, which is determined by the radius's rms value, $R_\alpha$ = 1.6782 fm.\\
Laha et.al. utilise screened Coulomb interaction based on Atomic Hulthen\cite{8} potential, given by
\begin{equation}
V_{H}(r)= V_o \frac{e^{-r/a}}{(1-e^{-r/a})}
\label{H}
\end{equation}
where $V_{o}$ is strength of the potential and $a$ is the  screening radius. The two parameters $V_o$ and $a$ are related by sommerfeld parameter \cite{9}\\

\begin{figure}[hbtp]
\caption{Coulomb interaction modeled using Atomic Hulthen and \textit{erf()} ansatzes}
\centering
\includegraphics[scale=0.3]{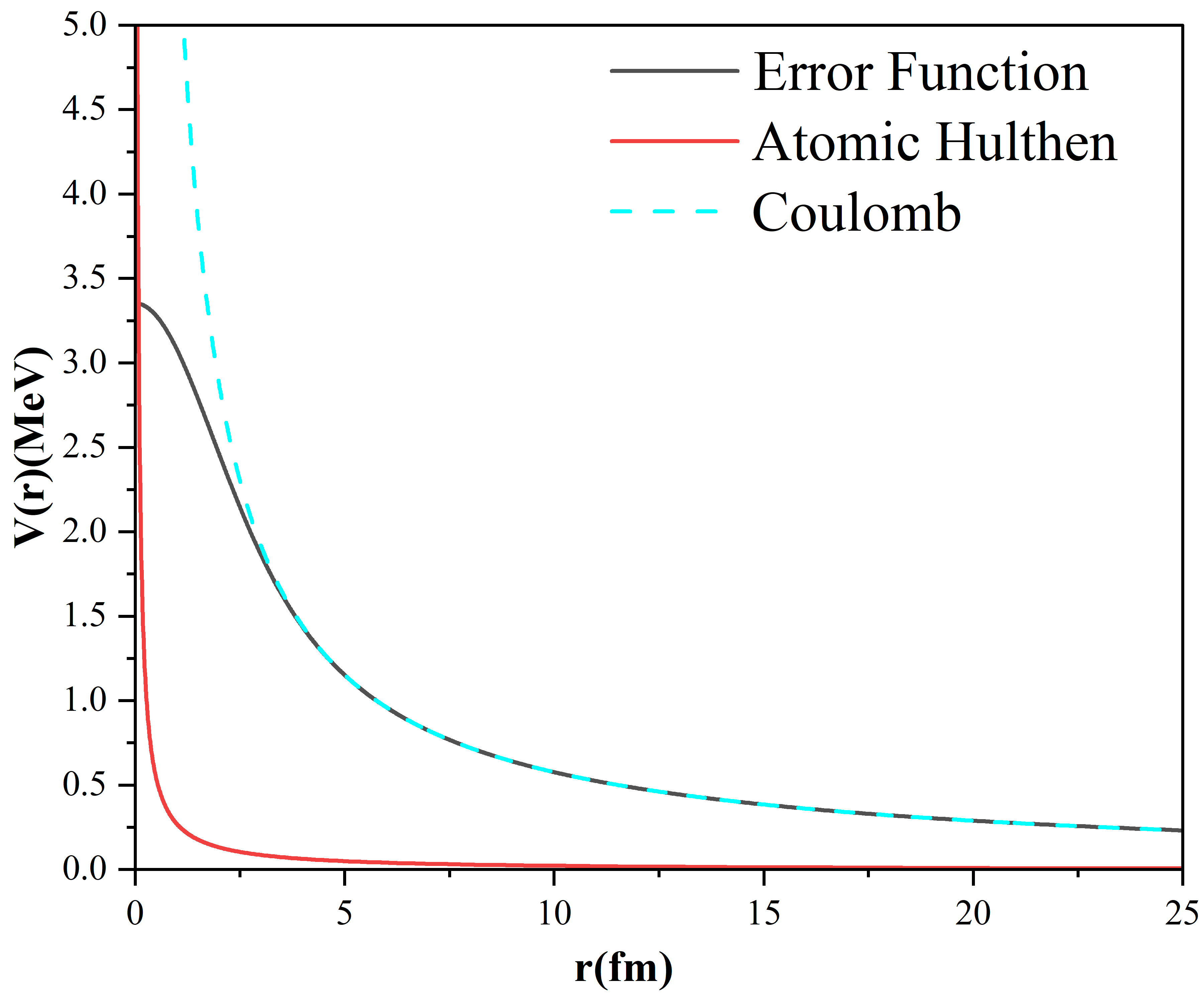}
\label{Fig1}
\end{figure}

For $\alpha - \alpha$, $Z_1 = Z_2 = 2$ , $\mu = \frac{m_{\alpha}}{2} = 1864.38525$ $\frac{MeV}{c^2}$ , 
$e^2 = 1.44 MeV fm$ and therefore $V_o a = 0.2758 fm^{-1}$. 
This ansatz is also plotted in Fig.\ref{Fig1}.
While the \textit{erf}-based screening potential has a finite value as r approaches 0 and needs to be abruptly cutoff for bigger values of r due to its long range behaviour, the Atomic Hulthen potential has an infinite value as r approaches 0.\\
Recently, Laha et.al.\cite{10}, have incorporated 
a separable non-local potential in place of the short range nuclear potential and obtained analytical solution of Schrodinger equation.
By combining Manning-Rosen and Yamaguchi potentials, they were able to describe the scattering of two charged particles and, using the Green's function approach, find regular and irregular solutions.
In essence, they have replaced Hulthen function with a more generic version by using Manning-Rosen potential as a model of interaction for screened and cut-off Coulomb interaction.\\
On the other hand, we have proposed a computational approach to construct inverse potentials\cite{11, 12} based on phase function method. The methodology involves utilizing Morse potential as zeroth reference \cite{11} and optimizing the model parameters by considering all available experimental scattering phase shifts as in inverse scattering theory. This is called as reference potential approach\cite{13,14}. This is akin to machine learning paradigm wherein one obtains the model from available data either using a global optimisation algorithm\cite{15} or neural networks\cite{16}.\\
Even while this works quite well for determining the inverse potentials for scattering when either of the two particles, projectile or target, is neutral, an additional Coulomb potential is required for charged systems. Previously, while studying $pp$ \cite{17}, $pd$ \cite{18}, $p-\alpha$ \cite{19} and $\alpha-\alpha$ \cite{20} systems, we have chosen to incorporate \textit{erf()} based Coulomb potential. In order to get accurate scattering phase shifts, we cut off the \textit{erf()} at certain distance $r_f$ abruptly, which is a major limitation of this approach. Recently, we have undertaken a study of $\alpha$- $\alpha$ scattering using various phenomenological potentials by utilising Atomic Hulthen as screened coulomb potential \cite{21}. It was observed that screening radius had to be varied for different $\ell$-channels so as to obtain good match with experimental data.\\
A further review of literature revealed an interesting approach suggested by Selg\cite{13,14}, where the molecular interaction potentials were obtained by considering a combination of two or three Morse functions. Here's a method for using the experimental data at hand to construct a reference potential for a one-dimensional quantum mechanical system. The key advantage of this approximation over a linear fit is the low number of analytically distinct components required to provide a decent match with the original potential across a broad distance range.\\
Because a two-particle problem can always be reduced to a one-particle problem in a spherically symmetric field, the one-dimensional inverse theory may be applied to two body systems. As a result, the primary goal of this study is to develop inverse potentials by examining elastic scattering of charged $\alpha$ particles with $^4He$ having orbital angular momentum $l = 0, 2, 4$ at energies up to $28 MeV$, utilizing a combination of two Morse potentials across two distinct regions. The conventional Morse function is the initial potential at short distances, while the second portion is an inverted Morse function to capture the barrier height owing to Coulomb interaction.\\
\textit{Methodology:}
Theoretically, constructing inverse potentials \cite{22} requires not only information regarding all the bound state energies $E_n (n=0,1,\ldots,N)$ along with their related normalization constants $C_n$ but also the scattering phase shifts for all scattering energies $E > 0$ ranging to infinity. 
Most often, phase shift data is available for only certain energies to within a limited range and hence, a rigorous solution of quantum mechanical inverse problem is extremely difficult to compute.\\
\textit{Reference Potential Approach(RPA):} Selg\cite{23} suggests reference potential approach for solving 1D quantum systems wherein a single Morse function\cite{24} or a combination of smoothly joined Morse potentials of the form
\begin{equation}
V_k^{RPA}(r) = V_k + D_k[e^{-2\alpha_k(r-r_k)}-2e^{-\alpha_k(r-r_k)}]~~~~~k = 1, 2, \ldots
\label{U}
\end{equation}
can be chosen as starting point to solve time-independent Schrodinger equation for its energy eigenvalues, scattering phase shifts, and also Jost function, from which one can obtain the inverse potential\cite{14}.\\ 
Here $D_k$'s are potential depths at  equilibrium distances $r_k$'s, and $\alpha_k$'s reflect shape parameter of Morse functions. $V_k$'s are constants added to total potential, whose importance shall be made clear later. These functions are smoothly joined at various boundary points $X_{k+1}$. \\
The number of distinct Morse-type components that may be added is almost unlimited. Naturally, the more components one includes, the better the match with the experimental data, but also the more challenging the analytical solution to the problem gets.
\\ Only two Morse components are utilized in this entire letter to prepare the reference potential.
The first of these is a regular Morse component $(k = 1, D_1 > 0)$, while the second is a reversed Morse component $(k = 2, D_2 < 0)$. The potential has a minimum $R_1$ in the range of $0 < R_1 < X$, which uses regular Morse component and a maximum $R_2 > X$ due to reversed Morse potential with a negative $D_2$ value. Despite the fact that this may seem unreasonable and irrational, the objective is that the height of this artificial hump approaches zero as the parameter $r_2$ goes to infinity. As a result, with a large enough $r_2$, the hump almost completely vanishes, yet the analytical strategy remains simple and flexible.
 For ensuring smoothness of potential at the boundary point $X$, in-between the two, the functions and their derivatives must be continuous at $X$. That is, 
\begin{eqnarray}
U_1(r)|_X = U_2(r)|_X \nonumber\\
\frac{dU_1(r)}{dr}\Big|_X = \frac{dU_2(r)}{dr}\Big|_X
\label{continuity}
\end{eqnarray}
Using these equations, two of the six parameters were determined as
\begin{equation}
D_1=\frac{(V_2-V_1)\alpha_2 f_2}{\alpha_2 f_3 f_2 -\alpha_1 f_4 f_1}\\
\label{D1}
\end{equation}
\begin{equation}
D_2=\frac{(V_1-V_2)\alpha_1 f_1}{\alpha_2 f_3 f_2 -\alpha_1 f_4 f_1}
\label{D1D2}
\end{equation}
where
\begin{eqnarray}\nonumber
f_1&=&e^{-2\alpha_1(r-r_1)}-e^{-\alpha_1(r-r_1)}\\
f_2&=&e^{-2\alpha_2(r-r_2)}-e^{-\alpha_2(r-r_2)}\nonumber \\
f_3&=&e^{-2\alpha_1(r-r_1)}-2e^{-\alpha_1(r-r_1)}\nonumber \\
f_4&=&e^{-2\alpha_2(r-r_2)}-2e^{-\alpha_2(r-r_2)}\nonumber 
\end{eqnarray}
Now, observing Eqs. \ref{D1} and \ref{D1D2}, one can see that not choosing $V_1$ and $V_2$ would have led to two homogeneous equations and determination of $D_1$ and $D_2$ wouldn't have been feasible. It is the difference $V_1-V_2$ that is important, and in fact, our results show that $V_2$ is always zero. 

While Selg\cite{25} obtains analytical expressions for energies and phase shifts by solving radial Schrodinger equation for molecular potentials, we take an alternative computational approach by employing phase function method. 
\\
\textit{Phase Function Method:} The second-order time-independent Schrodinger equation for a potential $V(r)$, 
\begin{equation}
\frac{\partial^2\psi(r)}{\partial r^2}+\frac{2m}{\hbar^2}[E-V(r)]\psi(r)=0
\label{1}
\end{equation}
is recast into a first order non-linear differential equation of Riccati type 
\begin{equation}
\frac{d\delta_l(k,r)}{dr}=-\frac{U(r)}{k}[\hat{j_l}(kr)cos( \delta_l(k,r))-\hat{\eta_l}(kr)sin(\delta_l(k,r))]^2
\label{5}
\end{equation}
where $k=\sqrt{2\mu E/{\hbar^2}}$, $U(r)=2\mu V(r)/\hbar^2$, $\hat{j_l}(kr)$ and $\hat{\eta_l}(kr)$ are the Riccati–Bessel and Riccati-Neumann functions, respectively. This gives phase shift information at various energies by directly taking potential as input.\\
For S($\ell = 0$), D($\ell=2$) and G($\ell=4$) states, the corresponding phase equations \cite{20} can be obtained by substituting the appropriate $\ell$th order Riccati-Bessel and Riccati-Neumann functions as follows:
\begin{description}
\item[Case 1: $\ell = 0$]
\begin{equation}
    \delta_0'(k,r)=-\frac{U(r)}{k}\sin^2(\delta_0 + kr)
\end{equation}
\item[Case 2: $\ell$ = 2]
\begin{eqnarray}\nonumber
\delta_2'(k,r) &=& -\frac{U(r)}{k}\bigg[-\sin{\left(\delta_2+ kr \right)}\\
&&-\frac{3 \cos{\left(\delta_2 + kr \right)}}{kr} + \frac{3 \sin{\left(\delta_2 + kr \right)}}{(kr)^2}\bigg]^2 
\end{eqnarray}
\item[Case 3: $\ell$ = 4]
\begin{eqnarray}\nonumber
\delta_4'(k,r)&=&-\frac{U(r)}{k}\bigg[\sin{\left(\delta_4 + kr \right)} \\ \nonumber
& & +  \frac{10 \cos{\left(\delta_4 + kr \right)}}{kr}-\frac{45 \sin{\left(\delta_4 + kr \right)}}{ (kr)^2}\\
& & +
\frac{105 \cos{\left(\delta_4 + kr \right)}}{ (kr)^3}+ \frac{105 \sin{\left(\delta_4 + kr \right)}}{ (kr)^4}\bigg]^2
\end{eqnarray}

\end{description}

These equations are solved using 5th order Runge-Kutta method by choosing the initial condition as $\delta_{\ell}(0, k) = 0$ and integrating to a large distance.
So, one can optimize the model parameters of chosen potential by iteratively solving the phase equations for various energies and minimizing a cost function. Typically, one can employ mean squared error, mean absolute percentage error(MAPE), $\chi^2$-error, etc., between computed and experimental scattering phase shifts, as cost function.\\

\begin{figure}[h!]
\centering
\includegraphics[scale=0.5]{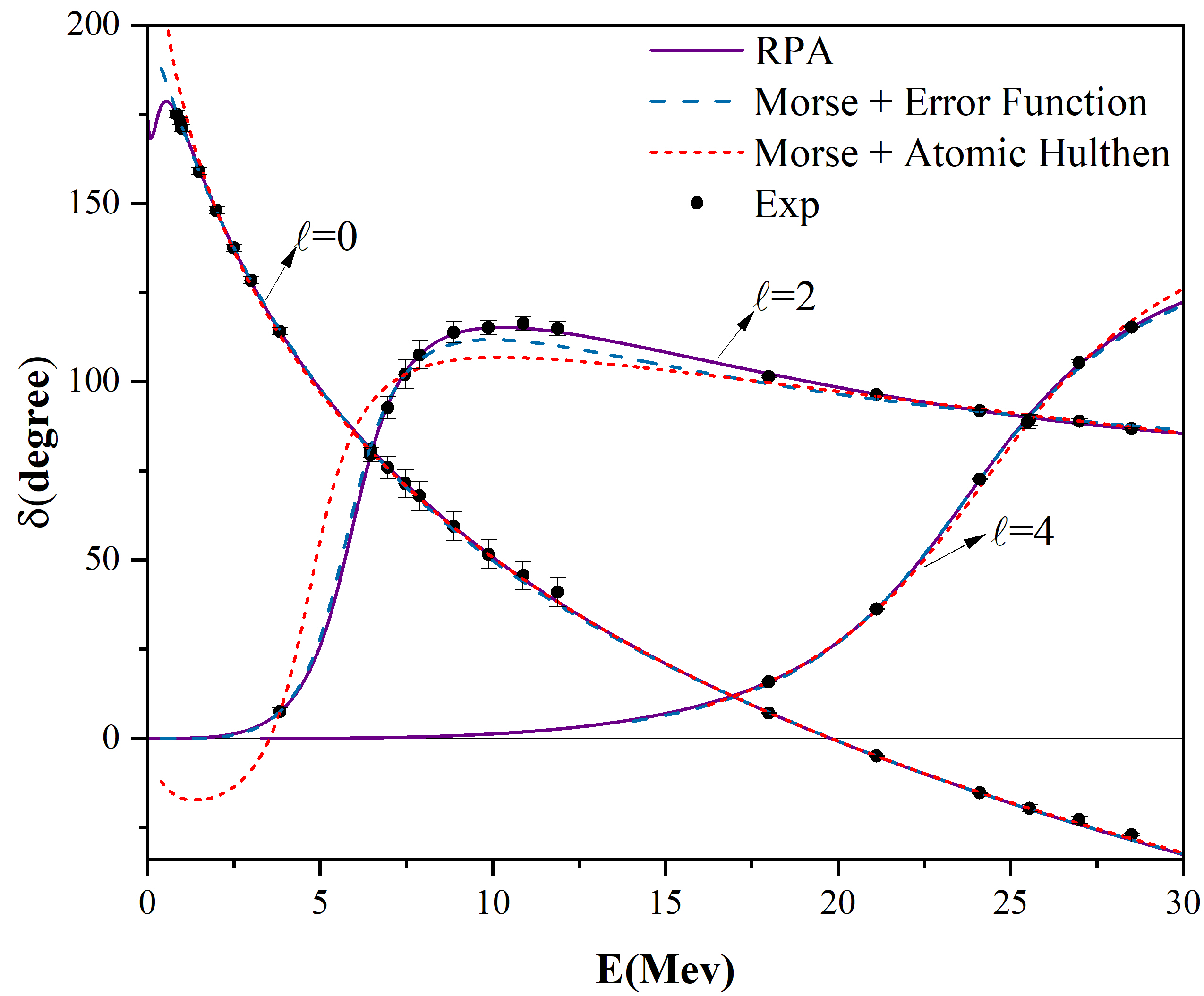}
\caption{$\alpha-\alpha$ scattering phase shifts for $\ell$ = 0. 2 and 4 channels obtained using various ansatze in comparison with experimental data}
\label{fig:2}
\end{figure}

\textit{Results and Discussion:}The updated experimental data for $\alpha-\alpha$ scattering by Anil et. al.,\cite{20} had included the Chein and Brown data \cite{26} for energies up to 25.5 MeV only. This has limited the available data points for $\ell = 4$ channel to only 4 energy values out of the 6 energies at which scattering phase shifts were given by \cite{26}. In our current analysis, we have opted to use a combination of two Morse potentials. The first part for a region, 0 $\leq$ r $\leq$ X, is a regular Morse with 4 parameters and the second part for larger distances, X $\leq$ r $\leq$ r$_f$, has an inverse Morse function with another 4 parameters. These two have to produce a smooth shaped potential that accounts for the observed scattering phase shifts. So, employing continuity conditions as given in Eqs. \ref{continuity}, one obtains the two parameters $D_1$ and $D_2$ in terms of the rest of the parameters as given by Eqs.\ref{D1} and \ref{D1D2}. This reduces the number of parameters required to 6 but one needs to fix point $X$ as well. So, observing that the relative difference $V_1-V_2$ appears in both Eqs.\ref{D1} and \ref{D1D2}, we have chosen to set $V_2 = 0 $. In fact, we have initially run the optimization routine by fixing different values of $X$ and observed that $V_2$ always comes out to be of the order of $10^{-8}$. So, in this way, we have ensured that only 6 parameters need to be determined by utilizing the 6 experimental data points that are available for $\ell=4$. Hence, we have chosen experimental data all the way up to 28 MeV, just below the threshold binding energy of alpha particles, for all the $\ell$ channels.\\
Other important physical considerations while optimizing the model parameters to obtain inverse potentials are 
\begin{enumerate}
\item The Coulomb barrier for the $\ell = 0$ potential needs to be high enough to ensure it has a pseudo-bound state with energy 0.0918 MeV. 
\item After accounting for centrifugal potential, the depth of the $\ell = 2$ potential must be lower than the $\ell = 0$ potential, and the Coulomb barrier height must be on the order of resonance energy, which has been measured to be approximately 3 MeV.
\item Similarly, depth of $\ell = 4$ potential must be less than that of $\ell = 2$ and Coulomb barrier height must be of the order of 12 MeV, which is close to the observed resonance energy for this channel.
\end{enumerate} 
 As a first step, we have reconstructed the inverse potentials for $\ell = 0,~2,~4$ channels of $\alpha-\alpha$ scattering by choosing single Morse potential with $V_1=0$ in Eq. \ref{U} as a model for nuclear interaction and considering Coulomb interaction as \textit{erf()} ansatz in Eqs.\ref{error} and Atomic Hulthen form in Eqs.\ref{H}, one at a time. The Mean Absolute Percentage Error(MAPE), between obtained and experimental values, has been minimized for optimizing the model parameters. The best model parameters obtained, along with MAPE, are given in Table \ref{Table1}. \\
\begin{table}
\centering
\caption{Parameters for nuclear interaction modeled using Morse function for various $\ell$-channels of $\alpha-\alpha$ scattering using two different ansatzes for Coulomb interaction}

\begin{tabular}{c|ccc|ccc}
\hline
Model & \multicolumn{3}{c|}{Morse + erf()} & \multicolumn{3}{c}{Morse + Atomic Hulthen} \\  
Parameters & \multicolumn{1}{c}{$\ell=0$} & \multicolumn{1}{c}{$\ell=2$} & $\ell=4$ & \multicolumn{1}{c}{$\ell=0$} & \multicolumn{1}{c}{$\ell=2$} & $\ell=4$ \\ \hline
$D_1$ & \multicolumn{1}{c}{11.1792} & \multicolumn{1}{c}{27.6925} & 182.7336 & \multicolumn{1}{c}{9.8096} & \multicolumn{1}{c}{26.4894} & 240.2966 \\
$\alpha_1$ & \multicolumn{1}{c}{0.5661} & \multicolumn{1}{c}{1.4539} & 1.2393 & \multicolumn{1}{c}{0.5318} & \multicolumn{1}{c}{1.5241} & 1.3572 \\ 
$r_1$ & \multicolumn{1}{c}{3.4684} & \multicolumn{1}{c}{1.8507} & 0.4768 & \multicolumn{1}{c}{3.6399} & \multicolumn{1}{c}{1.8745} & 0.3695 \\ 
$V_d$ & \multicolumn{1}{c}{-9.5506} & \multicolumn{1}{c}{-8.6002} & 3.0134 & \multicolumn{1}{c}{-9.0924} & \multicolumn{1}{c}{-8.8917} & 0.9819 \\ 
$V_{CB}$ & \multicolumn{1}{c}{0.3543} & \multicolumn{1}{c}{3.2087} & 10.1654 & \multicolumn{1}{c}{0.0915} & \multicolumn{1}{c}{2.6126} & 10.7846 \\ 
MAPE & \multicolumn{1}{c}{1.6324} & \multicolumn{1}{c}{1.6332} & 1.1988 & \multicolumn{1}{c}{1.9601} & \multicolumn{1}{c}{4.0609} & 0.5675 \\ \hline
\end{tabular}%

\label{Table1}
\end{table}
It should be mentioned that $r_f$ for \textit{erf()} function has been taken as 10 fm, 6.2 fm and 5 fm and screening radius $a$ for Atomic Hulthen has been chosen as 20 fm, 8 fm and 3.5 fm for $\ell$ = 0, 2 and 4 channels respectively. The scattering phase shifts obtained using these two ansatzes are shown in Fig \ref{fig:2}. One can observe that both these ansatzes are unable to correctly explain experimental data for $\ell$ = 2.

Then, we have optimized the model parameters for combination of two Morse functions as suggested by reference potential approach, by choosing to vary all 7 of them including $X$ value simultaneously, for $\ell = 0$ and $\ell = 2$. It was observed that value of $V_2$ is 0 in both the cases. So, our choice of $V_2 = 0 $ for $\ell = 4$ is well justified. These parameters were obtained by choosing final integration distance $r_f$ to be 40 fm. One can, in principle, take $r_f$ to be as large as required to ensure that potential goes to zero with distance. The final obtained parameters for inverse potentials are given in Table \ref{Table2}. 

\begin{table}[!ht]
 \centering   
    \caption{Parameters for $\alpha-\alpha$ scattering using combination of two Morse functions as reference potential}
    \begin{tabular}{cccc}
    \hline
        \textbf{Model Parameters} & \textbf{$\ell=0$} & \textbf{$\ell=2$} & \textbf{$\ell=4$} \\ \hline 
        \textbf{$V_1$} & 0.7915 & 7.0575 & 39.1351 \\ 
        \textbf{$\alpha_1$} & 0.5675 & 0.8019 & 0.6877 \\ 
        \textbf{$r_1$} & 3.5362 & 1.5679 & 0.8934 \\ 
        \textbf{$\alpha_2$} & 0.3629 & 2.3229 & 0.8301 \\ 
        \textbf{$r_2$} & 12.0073 & 4.6009 & 7.8639 \\ 
         
        \textbf{X} & 8.0000 & 4.4723 & 2.7694 \\

        \textbf{$D_1$} & 10.9153 & 29.1528 & 125.4702 \\ 
        \textbf{$D_2$} & 0.0932 & 1.8835 & 0.0045 \\ 
        \textbf{$V_d$} & -10.3173 & -4.5497 & 0.8501 \\ 
          \textbf{$V_{CB}$} & 0.0969 &	4.8802
           &11.1349 \\ 
           \textbf{MAPE} & 0.8 & 0.5 & 0.4 \\  \hline
    \end{tabular}
    \label{Table2}
    
\end{table}

\begin{figure*}
\includegraphics[scale=0.36]{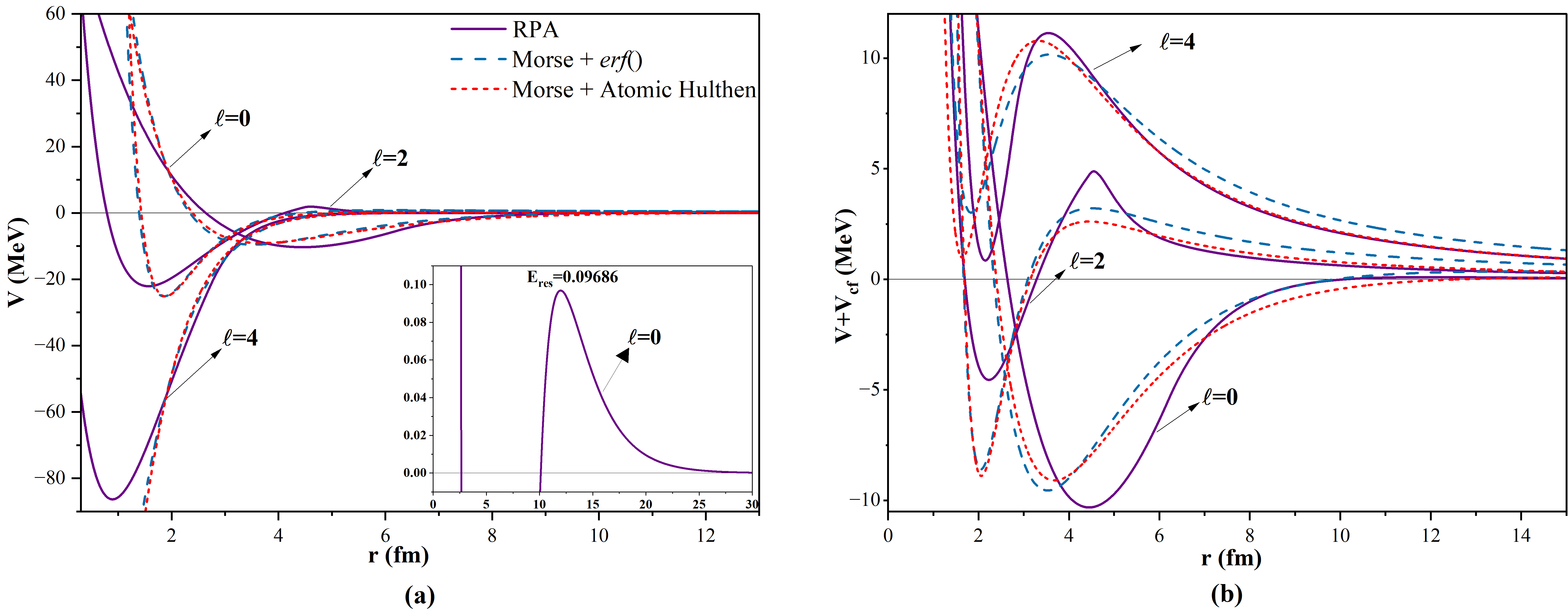}
\caption{\label{fig:3}Inverse potentials for $\ell=0,~2~\&~4$ using various ansatzes (a) without and (b) with centrifugal potential term added.}
\end{figure*}

\begin{figure}
 \centering
\includegraphics[angle=360,width = 0.5\textwidth]{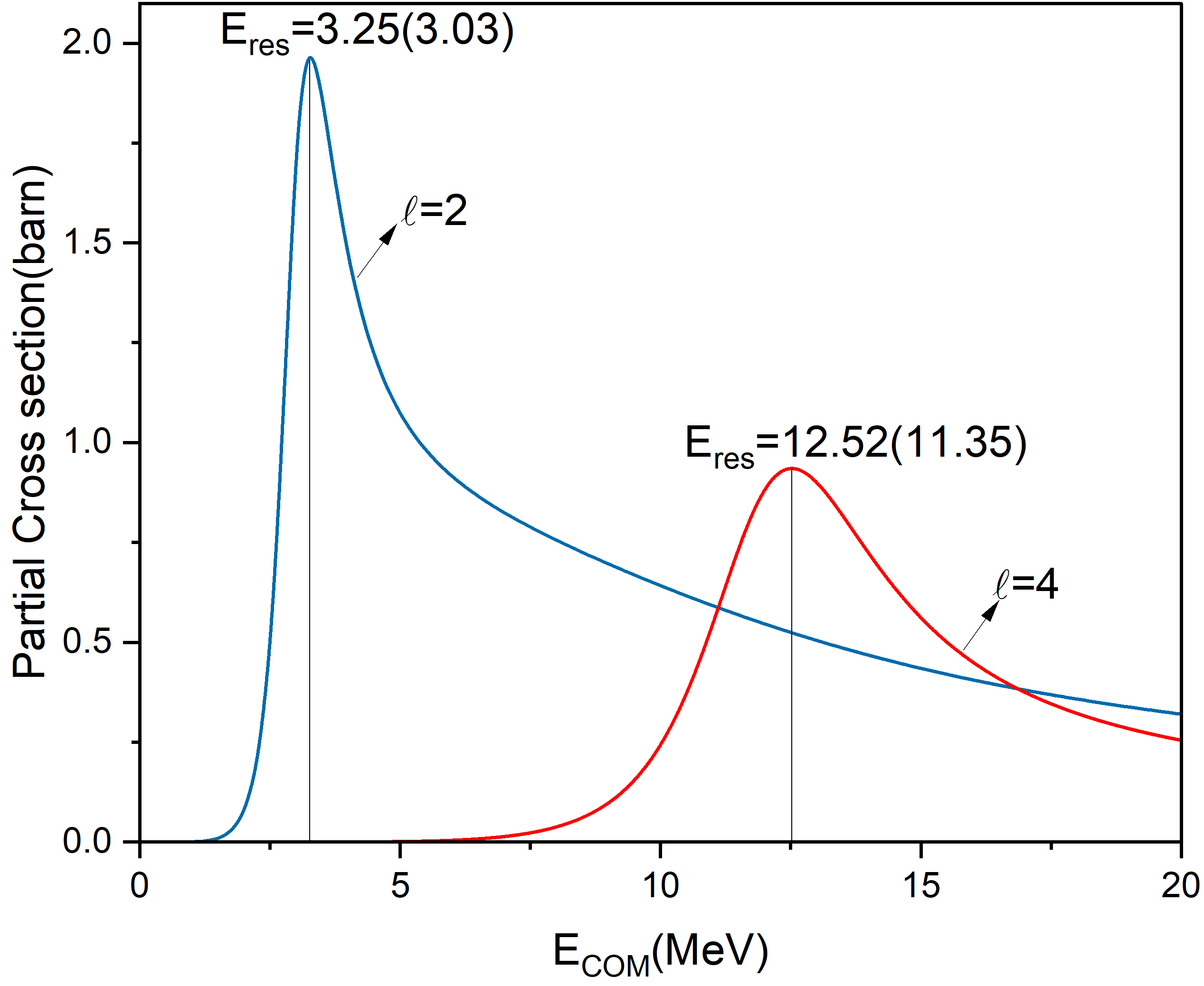}
\caption{Partial cross sections for $\ell = 2$ and $4$ for different center of mass energies}
\label{fig:5}
\end{figure}

The obtained scattering phase shifts, plotted in Fig. \ref{fig:2}, are observed to be matching very well with the experimental data. One can also compare the MAPE obtained using each of the ansatzes to appreciate the excellent performance of RPA as compared to screened Coulomb ansatzes. Especially, in case of $\ell$ = 2 channel,  depth of the potential $V_d$ after adding the centrifugal potential seems to reasonable lesser than the depth for $\ell = 0$ using RPA which is not the case for Morse + \textit{erf()} or Morse + Atomic Hulthen ansatzes. The constructed inverse potentials for all channels are plotted without considering centrifugal term $V_{cf} = \ell(\ell+1)/r^2$ in Fig. \ref{fig:3}(a) and by including it in \ref{fig:3}(b). The Coulomb barrier $V_{CB}$ heights using RPA can be observed to be always a little larger than what are obtained using the other two models. \\
The scattering phase shifts for $\ell = 0 $ become negative beyond 20 MeV reflecting that a repulsive core is present. While Morse with \textit{erf} and Atomic Hulthen models demonstrate that repulsion has a very hard core, RPA with two Morse potentials reveals that the height of the repulsive component is limited.  The inset of Fig. \ref{fig:3}(a) depicts the centrifugal barrier height of $\ell=0$ for RPA, and it is evident that the s-state is a bound state with resonance energy of $E_{res}=0.09686$.
make the S-state to be a pseudo-bound state. In case of $\ell = 2$, the scattering phase shifts initially increase very quickly until about 10 MeV and then show a slow decreasing trend for higher energies. Once again, a similar observation as for potentials of $\ell = 0 $ can be seen, with RPA having finite height in its repulsive part. Finally, in case of $\ell = 4$, even though the scattering phase shifts seems to be having an increasing trend, on close observation, one can see a small shift at higher energies towards peaking. In fact, at even higher energies than those considered in this work, one would find scattering phase shifts to peak and have a decreasing trend, as observed in case of $\ell = 2$ channel. So, one would expect that there would be a relatively small repulsive behavior that would set in even for $\ell = 4$ and is observed in case of RPA. \\
It is interesting to note that after adding centrifugal term, all the inverse potentials show a repulsion that is just beyond the radius of alpha particle. Further, the potential depths $V_d$ for various $\ell$-channels are distinctly decreasing in RPA as compared to the other two ansatzes for which $\ell=0$ and $\ell=2$ channels have depths that are very close. Even though all three models show Coulomb barrier heights $V_{CB}$ to be closer to resonances observed experimentally, it is only in RPA that they appear naturally without the imposition of any constraints. Remember that in the case of \textit{erf()} based Coulomb interaction, we had to limit the integration distance $r_f$ for various channels. Since the Coulomb potential is a long-range potential, the interaction potential obtained by using the error function is highly dependent on final integration point. Similarly, in the case of the Hulthen model, we had to adjust the screening radius for various channels to obtain good results. Both of these limitations have been overcome in RPA, in which the results obtained are independent of the integration distance.\\ 
\textit{Partial cross sections:}
The computed phase shifts are used to determine partial cross sections using 
\begin{equation}
\sigma_l=\frac{4\pi}{k^2}(2l+1)sin^2\delta_l(E)
\end{equation}
In case of $\ell$ = 0, there is a sharp resonance behavior very close to 0 and it appears almost like a delta function. As a result, we skipped plotting the partial cross section for this case.
In Figure \ref{fig:5}, the partial cross sections for $\ell = 2$ and $4$ channels are displayed and are seen to be in Breit-Wigner form. We obtain the resonance peaks(FWHM) for $\ell = 2$ and $4$ at 3.25(2.46) MeV and 12.52(5.91) MeV respectively, as compared to experimental values of 3.03(1.51) and 11.35(3.50) \cite{27}.\\ 
\\
\textit{Conclusion:} The reference potential approach is ideally suited for accurately describing the long range Coulomb interaction alongside the short range nuclear interaction for scattering involving charged particles. This procedure could overcome the limitations of previously employed ansatzes for Coulomb interaction, such as \textit{erf()} and 
 Atomic Hulthen. It will find applications in scattering studies of atomic, molecular, and nuclear domains. It would also be effective in studying nuclear and astrophysical reactions.\\
\\
\textit{Acknowledgement:}: A. Awasthi acknowledges financial support provided by Department of Science and Technology (DST), Government of India vide Grant No. DST/INSPIRE Fellowship/2020/IF200538.
   \\
 
\textbf{Author Declaration} The authors declare that they have no conflict of interest.


\begin{thebibliography}{100}
\bibitem{1} J. R. Taylor, A new rigorous approach to coulomb scattering , II  Nuovo Cimento B Series 11 \textbf{23}, 313 (1974)
\bibitem{2}E. P. Wigner and L. Eisenbud, Higher angular momenta and long range interaction in resonance reactions, Physical Review \textbf{72}, 29 (1947)..
\bibitem{3} M. Odsuren, K. Kato, G. Khuukhenkhuu, and S. Davaa, Scattering cross section for various potential systems, Nuclear Engineering and Technology \textbf{49}, 1006 (2017).
\bibitem{4} F. Calogero, \textit{ Variable Phase Approach to Potential Scattering by F Calogero} (Elsevier, 1967)
\bibitem{5}  B. Talukdar, D. Chattarji, and P. Banerjee, A generalized approach to the phase-amplitude method, Journal of Physics G: Nuclear Physics \textbf{3}, 813 (1977)
\bibitem{6}  S. Ali and A. Bodmer, Phenomenological $\alpha-\alpha$ potentials, Nuclear Physics \textbf{80}, 99 (1966).
\bibitem{7} B. Buck, H. Friedrich, and C. Wheatley, Local potential models for the scattering of complex nuclei, Nuclear Physics A \textbf{275}, 246 (1977).
\bibitem{8} J. Bhoi and U. Laha, Elastic scattering of light nuclei through a simple potential model, Physics of Atomic Nuclei \textbf{79}, 370 (2016).
\bibitem{9} D. Yakovlev, M. Beard, L. Gasques, and M. Wiescher, Simple analytic model for astrophysical s factors, Physical Review C \textbf{82}, 044609 (2010).
\bibitem{10}  P. Sahoo and U. Laha, Exact solutions of the schrodinger equation for manning-rosen plus yamaguchi potential, Physica Scripta \textbf{96}, 095302 (2021).
\bibitem{11} A. Khachi, L. Kumar, A. Awasthi, and O. S. K. S. Sastri, Inverse potentials for all channels of neutron-proton scattering using reference potential approach, Physica Scripta \textbf{98}, 095301 (2023)
\bibitem{12} O. Sastri, A. Khachi, and L. Kumar, An innovative approach to construct inverse potentials using variational monte-carlo and phase function method: Application to np and pp scattering, Brazilian Journal of Physics \textbf{52}, 58 (2022).
\bibitem{13} M. Selg, Formation and rotational-vibrational relaxation of diatomic rare gas excimers, Physica Scripta \textbf{52}, 287 (1995).
\bibitem{14} M. Selg, Reference potential approach to the quantum-mechanical inverse problem: I. calculation of phase shift and jost function (2005), arXiv:quant-ph/0506064[quant-ph].
\bibitem{15}A. Khachi, L. Kumar, M. R. G. Kumar, and O. S. K. S.Sastri, Deuteron structure and form factors: Using an inverse potential approach, Phys. Rev. C \textbf{107}, 064002 (2023).
\bibitem{16}  J. Keeble and A. Rios, Machine learning the deuteron, Physics Letters B \textbf{809}, 135743 (2020).
\bibitem{17} L. Kumar, A. Khachi, A. Sharma, and O. Sastri, P $\&$ d inverse potentials for proton-proton scattering, in Proceedings of the DAE Symp. on Nucl. Phys, Vol. 66 (2022)p. 579
\bibitem{18} S. Awasthi and O. S. K. S. Sastri, Real and Imaginary Phase Shifts for Nucleon-Deuteron Scattering using Phase Function Method, (2023), arXiv:2304.10478
[nucl-th]

\bibitem{19} L. Kumar, S. Awasthi, A. Khachi, and O. S. K. S. Sastri, Phase Shift Analysis of Light Nucleon-Nucleus Elastic Scattering using Reference Potential Approach, (2022), arXiv:2209.00951 [nucl-th]..
\bibitem{20} A. Khachi, L. Kumar, and O. Sastri, Alpha–alpha scattering potentials for various $\ell$-channels using phase function method, Physics of Atomic Nuclei \textbf{85}, 382 (2022).
\bibitem{21} A. Awasthi and O. S. K. S. Sastri, Comparative Study of alpha-alpha interaction potentials constructed using various phenomenological models, (2023), arXiv:2307.13207 [nuclth].
\bibitem{22} V. A. Marchenko, Certain problems in the theory of second-order differential operators, in \textit{Doklady Akad. Nauk SSSR}, Vol. 72 (1950) pp. 457–460
\bibitem{23} M. Selg, Numerically complemented analytic method for solving the time-independent one-dimensional schrodinger equation, Physical Review E \textbf{64}, 056701 (2001).
\bibitem{24}M. Selg, A practical method for constructing a reflectionless potential with a given energy spectrum, Proceedings of the Estonian Academy of Sciences \textbf{65}, 358 (2016).
\bibitem{25} M. Selg, Visualization of rigorous sum rules for franck–condon factors: spectroscopic applications to xe2, Journal of Molecular Spectroscopy \textbf{220}, 187 (2003).
\bibitem{26}W. S. Chien and R. E. Brown, Study of the $\alpha-\alpha$ system below 15 mev (c.m.), Phys. Rev. C \textbf{10}, 1767 (1974).
\bibitem{27}D. Tilley, J. Kelley, J. Godwin, D. Millener, J. Purcell, C. Sheu, and H. Weller, Energy levels of light nuclei a=8, 9, 10, Nuclear Physics A \textbf{745}, 155 (2004)


\end{thebibliography}
\end{document}